\documentclass[12pt]{iopart}
\usepackage{graphicx}

\begin{document}

\newcommand{\PrOsSb}{PrOs$_4$Sb$_{12}$}
\newcommand{\LaOsSb}{LaOs$_4$Sb$_{12}$}

\title{Crystalline electric field effects in the electrical resistivity of \PrOsSb}

\author{N. A. Frederick and M. B. Maple}

\address{Department of Physics and Institute for Pure and
Applied Physical Sciences, University of California, San Diego, La
Jolla, CA 92093 USA}

\date{\today}

\begin{abstract}

The temperature $T$ and magnetic field $H$ dependencies of the
electrical resistivity $\rho$ of the recently discovered heavy
fermion superconductor \PrOsSb{} have features that are associated
with the splitting of the Pr$^{3+}$ Hund's rule multiplet by the
crystalline electric field (CEF). These features are apparently
due to magnetic exchange and aspherical Coulomb scattering from
the thermally populated CEF-split Pr$^{3+}$ energy levels. The
$\rho(T)$ data in zero magnetic field can be described well by
calculations based on CEF theory for various ratios of magnetic
exchange and aspherical Coulomb scattering, and yield CEF
parameters that are qualitatively consistent with those previously
derived from magnetic susceptibility, specific heat, and inelastic
neutron scattering measurements.  Calculated $\rho(H)$ isotherms
for a $\Gamma_{3}$ ground state qualitatively account for the
`dome-shaped' feature in the measured $\rho(H)$ isotherms.

\end{abstract}


\section{Introduction}

The filled skutterudite compound \PrOsSb{} was recently discovered
to be the first Pr-based heavy fermion superconductor, with a
superconducting transition temperature $T_\mathrm{c} = 1.85$ K and
an effective mass $m^{*} \approx 50~m_\mathrm{e}$, where $m_{e}$
is the mass of the free electron \cite{Maple01,Bauer02a}. Features
observed in the magnetic susceptibility $\chi(T)$, specific heat
$C(T)$, and inelastic neutron scattering (INS) \cite{Maple02a} all
indicate that the Hund's rule $J = 4$ multiplet of the Pr$^{3+}$
ion is split by the crystalline electric field (CEF).  An analysis
of these data within the context of a cubic CEF using the theory
of Lea, Leask and Wolf (LLW) yielded a $\Gamma_{3}$ nonmagnetic
doublet ground state, a $\Gamma_{5}$ triplet first excited state
at $\sim 10$ K, and higher $\Gamma_{4}$ triplet and $\Gamma_{1}$
singlet excited states at $\sim 130$ K and $\sim 313$ K,
respectively. A `roll-off' near $\sim 8$ K with electrical
resistivity $\rho(T)$ in zero magnetic field was attributed to a
decrease in spin or charge dependent scattering from the low-lying
$\Gamma_{5}$ excited state whose population decreases with
decreasing temperature, while features in $\rho(H,T)$ at low
temperatures $< 1.5$ K and high fields $> 4.5$ T were identified
with a high field ordered phase (HFOP) of magnetic or quadrupolar
character \cite{Maple02a,Ho01}.

In a lanthanide compound, the conduction electrons interact with
the localized $4$f electrons through magnetic exchange and the
direct Coulomb interaction.  For lanthanide ions whose $4$f
electrons are hybridized with conduction electrons, the exchange
interaction can be negative, which favors screening of the
magnetic moments of the $4$f electrons by the conduction electron
spins.  This leads to contributions to the electrical resistivity
that increase as $log~T$ with decreasing temperature that are
associated with the Kondo effect.  Since no Kondo-like
contribution has been found in $\rho(T)$, we disregard $4$f
conduction electron hybridization and treat the $4$f electrons in
the ionic limit, as a first approximation.  When the degeneracy of
the Hund's rule multiplet of a lanthanide ion is lifted in a
crystalline electric field, the change of the resulting lanthanide
energy level populations with temperature introduces
temperature-dependent features in the electrical resistivity.
These features are more subtle than those seen in properties such
as magnetic susceptibility and specific heat, but can still be
described based on the CEF Hamiltonian of Lea, Leask, and Wolf
(LLW) \cite{Lea62}.

This paper presents calculations of the electrical resistivity due
to magnetic exchange and aspherical Coulomb scattering associated
with the CEF splitting of the Pr$^{3+}$ Hund's rule multiplet in
\PrOsSb. Electrical resistivity $\rho(T)$ data in zero magnetic
field are directly fit with the calculated equations, providing
parameters that are used to calculate $\rho(H)$ isotherms.  It is
shown that a $\Gamma_{3}$ ground state can qualitatively account
for the `dome-shaped' features that have been observed in the
measured $\rho(H)$ isotherms.

\section{Calculation of the electrical resistivity in the crystalline electric field}

The effects on electrical resistivity due to magnetic exchange
\cite{Hirst67,Anderson74} and aspherical Coulomb \cite{Elliott54}
scattering have been separately considered many times in the past;
often one effect was neglected over the other.  The present
calculations follow the procedure of Fisk and Johnston
\cite{Fisk77}, where it was demonstrated that both contributions
are important for describing the temperature dependence of the
electrical resistivity for the compound PrB$_{6}$ and, in
addition, the relevant equations were presented in a form
appropriate to the Stevens operator equivalent formulation of the
LLW crystal field Hamiltonian \cite{Stevens52,Hutchings64}. In
this notation \cite{Anderson74}, the total contribution to the
electrical resistivity from CEF effects, $\rho_\mathrm{CEF}$, is:
\begin{equation}
\rho_\mathrm{CEF} =
\rho_{0}[r$Tr$(PQ^\mathrm{M})+(1-r)$Tr$(PQ^\mathrm{A})],
\label{rhoCEF}
\end{equation}
where $r$ is a coefficient representing the ratio of the magnetic
exchange term to the aspherical Coulomb scattering term.  The
temperature dependent matrix, $P_{ij}$, is common to both terms:
\begin{equation}
P_{ij} = \frac{e^{-\beta{}E_{i}}}{\sum_{k}e^{-\beta{}E_{k}}}\cdot
\frac{\beta(E_{i}-E_{j})}{1-e^{-\beta(E_{i}-E_{j})}}, \label{Pij}
\end{equation}
where $E_{i}$ are the eigenvalues of the CEF eigenstates and
$\beta = 1/k_\mathrm{B}T$.  The $Q_{ij}^\mathrm{M}$ matrix
represents magnetic exchange scattering, and the
$Q_{ij}^\mathrm{A}$ matrix is associated with aspherical Coulomb
scattering due to the quadrupolar charge distribution of the
Pr$^{3+}$ ion. The $Q_{ij}$ matrices are:
\begin{equation}
Q_{ij}^\mathrm{M} = |\langle{}i|J_{z}|j\rangle|^{2} +
\frac{1}{2}|\langle{}i|J_{+}|j\rangle|^{2} +
\frac{1}{2}|\langle{}i|J_{-}|j\rangle|^{2},
\end{equation}
\begin{equation}
Q_{ij}^\mathrm{A} =
\sum_{m=-2}^{+2}|\langle{}i|y_{2}^{m}|j\rangle|^{2},
\end{equation}
where the $|i\rangle$'s are the CEF eigenstates, and the
$y_{2}^{m}$'s are the operator equivalents of the spherical
harmonics for $L = 2$ (i.e., quadrupolar terms), and are given
elsewhere \cite{Fulde86}. The $Q_{ij}$ matrices are also
normalized to each other \cite{Fisk77}, such that
\begin{equation}
\sum_{i,j}Q_{ij}^\mathrm{M} = \sum_{i,j}Q_{ij}^\mathrm{A} =
(2J+1)J(J+1) = 180~(\mathrm{for}~J = 4).
\end{equation}

\begin{figure}[t]
\begin{center}
\includegraphics[width=12cm]{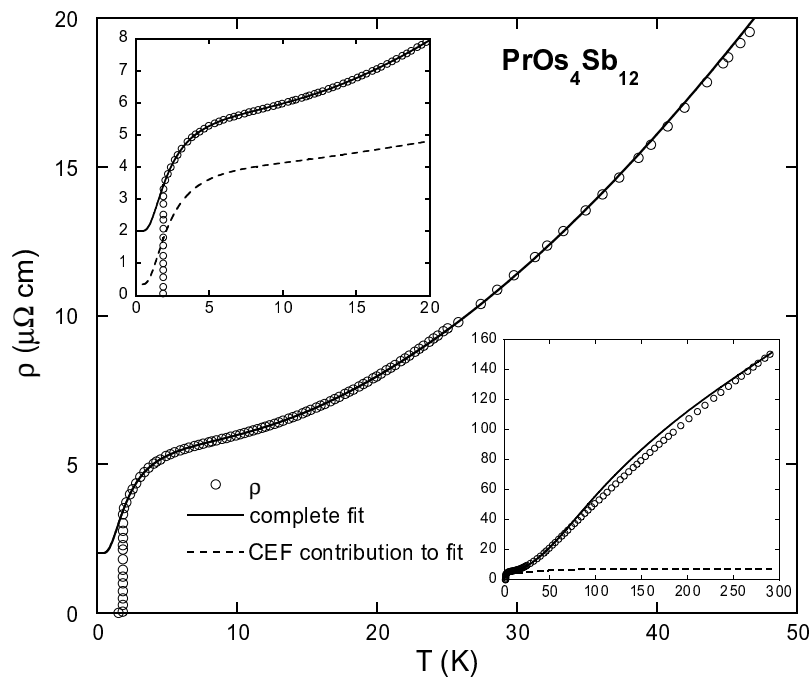}
\end{center}
\caption{\label{fit} Electrical resistivity $\rho$ vs temperature
$T$ between $1$ K and $50$ K for a single crystal of \PrOsSb.  The
solid line is a fit of the function described in the text
(Eq.~\ref{rhototal}), consisting of contributions from the
scattering of electrons by impurities, phonons (represented by the
scaled phonon contribution to the resistivity of \LaOsSb), and
magnetic exchange and aspherical Coulomb interactions. Upper
inset: $\rho$ vs $T$ between $1$ K and $20$ K for the same single
crystal of \PrOsSb.  The dashed line is the CEF contribution to
the fit represented by the solid line.  Lower inset: $\rho(T)$ of
\PrOsSb{} between $1$ K and $300$ K.  The CEF contribution (dashed
line) is nearly negligible at higher temperatures.}
\end{figure}

Using Matthieson's rule, the electrical resistivity is separated
into impurity, phonon, and CEF contributions:
\begin{equation}
\rho = \rho_\mathrm{imp} + A\rho_\mathrm{La} + \rho_{CEF}
\label{rhototal}
\end{equation}
where $\rho_\mathrm{imp}$ is the impurity scattering term,
$\rho_\mathrm{La}$ is the lattice term of the previously measured
isostructural compound without f-electrons, \LaOsSb{}
\cite{Maple02b}, and $\rho_{CEF}$ is given by Eq.(\ref{rhoCEF}).
The constant $A$ was used to scale the lattice contributions of
$\rho(T)$ of \LaOsSb{} and \PrOsSb{} to one another, assuming they
have the same temperature dependence, and to account for
uncertainties in the geometrical factor due to the irregular shape
of the crystals and/or microcracks in the crystals. For
$\rho_\mathrm{CEF}$, the energies $E_{i}$ can be expressed in
terms of the parameters $x$ and $W$ in accordance with the LLW
formalism \cite{Lea62}. In the absence of a magnetic field, the
eigenstates $|i\rangle$ are the same for all values of $x$ and $W$
for $J = 4$ in a cubic CEF. Additional terms due to tetrahedral
symmetry in the crystal field Hamiltonian \cite{Takegahara01} were
not included in this analysis in order to keep the calculations
consistent with those previously made on these samples.
Eq.~(\ref{rhototal}) was fit to the original data from
Ref.~\cite{Bauer02a} between $1.9$ K and $20$ K, assuming a
$\Gamma_{3}$ doublet ground state, with results shown in
Fig.~\ref{fit}. The best fit to the data yields an impurity
scattering term $\rho_\mathrm{imp} = 1.67~\mu\Omega$ cm, a scaling
factor for the \LaOsSb{} lattice $A = 0.21$, a CEF scaling factor
$\rho_{0} = 0.385~\mu\Omega$ cm, a CEF scattering mechanism ratio
$r = 0.46$, and LLW values of $x = -0.7225$ and $W = -2.97$, which
results in the first excited state $\Gamma_{5}$ triplet lying
$\sim 5$ K above the $\Gamma_{3}$ doublet ground state.

It is notable that the fit of Eq.~\ref{rhototal} to the zero field
$\rho(T)$ data between $T_\mathrm{c}$ and $20$ K is excellent
(upper inset of Fig.~\ref{fit}), and the agreement between the fit
and the data at high temperatures $> 50$ K is qualitatively good
(lower inset of Fig.~\ref{fit}). Since the magnetic exchange and
aspherical Coulomb scattering contributions to the CEF are nearly
temperature independent above $\sim 75$ K, the discrepancy is most
likely due to the \LaOsSb{} lattice term not being an accurate
representation of the \PrOsSb{} lattice term, which is reasonable
considering the anomalous resistivities of other La-based
compounds compared to their Y and Lu-based counterparts without
f-electrons \cite{Slebarski85}. Unfortunately, single crystals of
these YOs$_{4}$Sb$_{12}$ or LuOs$_{4}$Sb$_{12}$ compounds, which
may provide more accurate MOs$_{4}$Sb$_{12}$ filled skutterudite
lattice resistivities, are not available. The data were also
fitted with the CEF scattering mechanism ratio $r$ fixed at
intervals of $0.25$, and the results are presented in Table
\ref{fittable}. All values of $r$ resulted in satisfactory fits of
the $\rho(T)$ data at low temperatures, with the value of the
splitting between $\Gamma_{3}$ and $\Gamma_{5}$, $\Delta_{35}$,
increasing from $3.3$ K for $r = 1$ (magnetic exchange scattering
only) to $6$ K for $r = 0$ (aspherical Coulomb scattering only).
The values for $r = 0$ are closer to the values determined from
$\chi(T)$, $C(T)$, and INS measurements.  It is possible that a
better determination of the lattice term could result in a more
definitive result for the value of $r$.  While it is unlikely that
magnetic exchange scattering is completely unimportant, it does
appear that aspherical Coulomb scattering contributes considerably
to the electrical resistivity of \PrOsSb.

\begin{table}[t]
\caption{Parameters resulting from a fit of Eq.~\ref{rhototal} to
the data in Fig.~\ref{fit}.  The value of the CEF scattering ratio
$r$ was fixed at various values to determine how the other
parameters would change.  The larger the aspherical Coulomb
scattering contribution (smaller $r$), the closer the value of
$\Delta_{35}$ to values previously calculated.}
\begin{minipage}{\linewidth}\begin{indented}\item{}
\vspace{1cm}
\begin{tabular}{|l||c|c|c|c|c|c|}
    \hline
 \multicolumn{1}{|c||}{$r$} & $\rho_\mathrm{imp}$ & $A$ & $\rho_{0}$ & $x$ & $W$ & $\Delta_{35}$ \\
 & ($\mu\Omega$ cm) & & ($\mu\Omega$ cm) & & (K) & (K) \\ \hline
0 & 1.88 & 0.204 & 0.373 & $-0.716$ & $-2.52$ & 5.99 \\ \hline
0.25 & 1.80 & 0.207 & 0.378 & $-0.720$ & $-2.78$ & 5.33 \\ \hline
0.5 & 1.64 & 0.208 & 0.387 & $-0.723$ & $-3.00$ & 4.73 \\ \hline
0.75 & 1.33 & 0.206 & 0.406 & $-0.726$ & $-3.18$ & 3.93 \\ \hline
1 & 0.86 & 0.200 & 0.435 & $-0.728$ & $-3.30$ & 3.33 \\ \hline
\end{tabular}
\end{indented}\end{minipage}
\label{fittable}
\end{table}

\section{The high field ordered phase in \PrOsSb}

Measurements of the electrical resistivity, specific heat, and
thermal expansion of \PrOsSb{} at low temperatures $< 1.5$ K and
high magnetic fields $> 4.5$ T reveal features that appear to be
associated with an ordered phase, one that may be quadrupolar in
nature \cite{Maple02a,Ho01,Ho03,Vollmer03,Oeschler03}. In
electrical resistivity, these features are manifested as
`roll-offs' in $\rho(T)$ and large `domes' in $\rho(H)$
\cite{Maple02a,Ho03,Maple03}. The `domes' in $\rho(H)$ are
especially intriguing, as it is expected that scattering should be
lower in an ordered state.  It has been suggested that the
crossing of the CEF energy levels in magnetic fields are related
to the high field ordered phase \cite{Vollmer03,Maple03} and can
contribute to the increase of scattering in this region. Recent
neutron diffraction experiments have also suggested the existence
of an antiferro-quadrupolar ordered state that is associated with
magnetic field-induced energy level crossings for a $\Gamma_{1}$
ground state \cite{Kohgi03}. To check this hypothesis for
electrical resistivity, $\rho$ was calculated as a function of $H$
in order to qualitatively observe the shape of the CEF
contribution to the electrical resistivity.

\begin{figure}[t]
\begin{center}
\includegraphics[width=12cm]{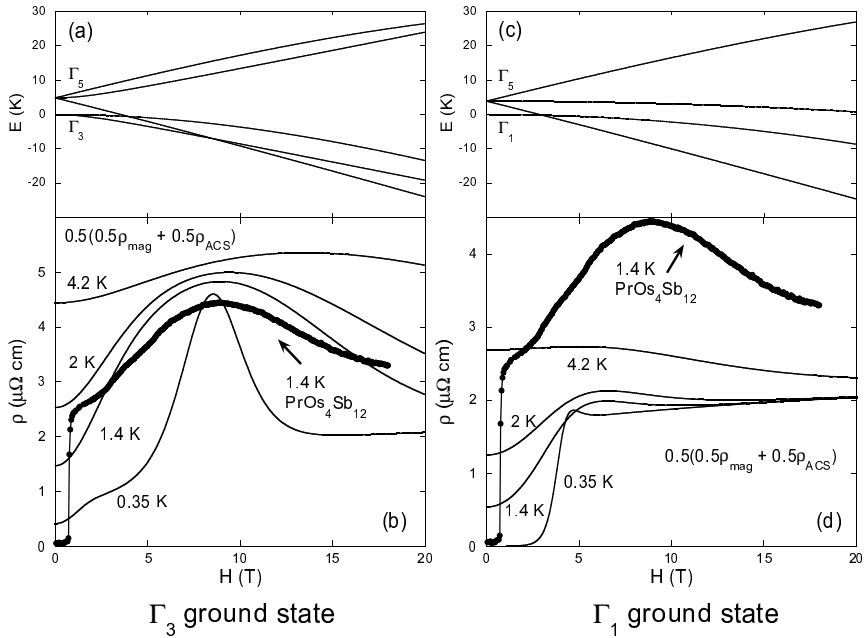}
\end{center}
\caption{\label{RvsH} (a) Low temperature energy level splittings
$E$ as a function of magnetic field $H$ for the LLW parameters $x
= -0.723$ and $W = -3.00$. (b) Calculated electrical resistivity
$\rho$ vs magnetic field $H$ for the energy splittings in (a).  It
was assumed that magnetic exchange and aspherical Coulomb
scattering contributed equally to the CEF resistivity ($r = 0.5$
in Eq.~\ref{rhoCEF}). The calculated lines were also scaled by
$\rho_{0} = 0.5$ in order to bring the curves to the same scale as
those measured experimentally.  For comparison, the dashed line is
the measured $\rho$ vs $H$ for \PrOsSb{} at $1.4$ K \cite{Ho03}.
(c) $E$ vs $H$ for the LLW parameters $x = 0.508$ and $W = 2.46$.
(d) $\rho$ vs $H$ for the energy splittings in (c), and similar
constraints as (b).}
\end{figure}

Shown in Fig.~\ref{RvsH}(a) are the energy level splittings as a
function of magnetic field ($H~\|~(001)$) for the $\Gamma_{3}$
doublet ground state and $\Gamma_{5}$ triplet first excited state,
using the LLW values of $x$ and $W$ derived from the fit to the
electrical resistivity of \PrOsSb{} with the CEF scattering
mechanism ratio $r$ chosen to be $0.5$. This value of $r$ was used
because it is consistent with the value determined by Fisk and
Johnston for PrB$_{6}$ \cite{Fisk77} as well as being a compromise
between aspherical Coulomb scattering, which results in a
$\Delta_{35}$ that is closer to values estimated from other
analyses ($\chi(T), C(T)$, INS), and magnetic exchange scattering,
which is more universal in lanthanide compounds.  The
corresponding plot of $\rho$ vs $H$ for various temperatures is
shown in Fig.~\ref{RvsH}(b). The parameter $\rho_{0}$ was chosen
so the calculated $\rho(H)$ curves were on the same scale as those
measured experimentally.  A dome shape, which sharpens
considerably at lower temperatures, can clearly be seen in the
graph. A plot of $\rho$ vs $H$ for \PrOsSb{} at $1.4$ K
\cite{Ho03} is presented for comparison to the theoretical curves.
The maximum of the dome corresponds to the the magnetic field at
which the ground state changes from one of the $\Gamma_{3}$ states
to one of the $\Gamma_{5}$ states. The dome structure can be
understood by taking the limit of the equation for the $P_{ij}$
terms, Eq.~\ref{Pij}, as the lowest excited state $E_{i} \to 0$,
assuming $E_{j} = 0$ is the ground state, and the other energy
levels are large enough to be ignored in the sum in the
denominator. Eq.~\ref{Pij} then becomes
\begin{equation}
\lim_{E_{i} \to 0} P_{ij} \approx \frac{\beta{}E_{i}}{2\sinh
\beta{}E_{i}}~\mathrm{for}~E_{j} = 0.
\end{equation}
The shape of this equation as a function of energy is a dome
centered on $E_{i} = 0$ that sharpens considerably as the
temperature is lowered.  Since $E_{i}$ is nearly proportional to
$H$ (Fig.~\ref{RvsH}(a)), domes also appear in $\rho(H)$.  These
domes are present in both magnetic exchange and aspherical Coulomb
scattering, as there are $\Gamma_{3}\to{}\Gamma_{5}$ transitions
in both mechanisms, which unfortunately contributes to uncertainty
in determining the relative importance between the two types of
scattering.

Similar to CEF fits to magnetic susceptibility $\chi(T)$, it is
also possible to fit $\rho(T)$ for \PrOsSb{} at low-temperatures
$T \leq 20$ K with an energy level scheme that has a $\Gamma_{1}$
singlet ground state and a $\Gamma_{5}$ triplet first excited
state.  The fit resulted in similar impurity and lattice fitting
parameters as for a $\Gamma_{3}$ ground state, with LLW values of
$x = 0.508$ and $W = 2.46$.  The parameters from the
zero-temperature fit were used to calculate the energy level
splittings as a function of magnetic field, as shown in
Fig.~\ref{RvsH}(c), with the corresponding plot of $\rho$ vs $H$
shown in Fig.~\ref{RvsH}(d). It is immediately apparent that the
dome shapes in $\rho(H)$ due to CEF scattering are less pronounced
and are not in agreement with the data when the ground state is
$\Gamma_{1}$, even though the fits to $\rho(T, H=0)$ are nearly
identical. A qualitative reason for this is that
$\Gamma_{1}\to{}\Gamma_{5}$ transitions only occur in aspherical
Coulomb scattering, and the energy levels cross at a much lower
field than for the $\Gamma_{3}$ ground state situation.

\section{Summary}

Features in the electrical resistivity of \PrOsSb{} can be
explained in terms of effects arising from the temperature- and
magnetic field-dependent populations of crystalline electric field
energy levels.  Fits to the $\rho(T)$ data were accomplished by a
combination of magnetic exchange and aspherical Coulomb scattering
in the context of a $\Gamma_{3}$ ground state.  The parameters
resulting from this fit were used to calculate isotherms of
$\rho(H)$ that agree qualitatively well with isotherms previously
measured.  While it was possible to fit the $\rho(T)$ data well
with a Pr$^{3+}$ energy level scheme with a $\Gamma_{1}$ ground
state, when $\rho(H)$ isotherms were calculated using CEF
parameters for this case, they were not in good agreement with the
measured $\rho(H)$ isotherms. It is evident that the high field
ordered phase that is also observed in high-field measurements of
$C(T)$ and $\alpha(T)$ is closely related to the ground state
crossover in the energy level-magnetic field phase diagram.

\section*{Acknowledgements}

We would like to thank P. Allenspach, V. S. Zapf, and P.-C. Ho for
useful discussions. This research was supported by the U.S.
Department of Energy Grant No.~DE-FG03-86ER-45230, the U.S.
National Science Foundation Grant No.~DMR-00-72125, and the NEDO
International Joint Research Program.

\end{document}